\documentclass[10pts,showpacs,preprint,aps]{revtex4}
\usepackage{graphicx}
\usepackage{dcolumn}
\usepackage{bm}
\usepackage{amssymb}
\usepackage{amsmath}
\begin{document}
\setcounter{page}{1}
\vskip 2cm
\title
{The Komar mass function in the de-Rham-Gabadadze-Tolley non-linear theory of massive gravity}
\author
{Ivan Arraut}
\affiliation{Department of Physics, Osaka University, Toyonaka, Osaka 560-0043, Japan}

\begin{abstract}
I derive the Komar mass/function for the Schwarszchild de-Sitter (S-dS) black-hole inside the dRGT non-linear theory of massive gravity by taking the usual notion of time-like Killing vector in unitary gauge. The dRGT Komar function depends on the dynamics of the St\"uckelberg fields through the gauge transformation function. It goes to the standard value obtained in General Relativity (GR) if the spatial derivative of the gauge function vanishes. In such a case, the (gauge) function corresponds to the usual notion of time as in GR.
\end{abstract}
\maketitle 
\section{Introduction}
If we want to modify gravity in order to reproduce the effects of dark energy, one possibility is to modify the Einstein-Hilbert action, such that non-derivative terms interactions are introduced in addition with the standard second derivative terms. On the other hand, if we want to reproduce dark matter effects, first derivative terms in the action should appear \cite{Deff,My paper}.
The dRGT theory is able reproduce dark energy effects inside a ghost-free formulation of the non-linear massive gravity \cite{deRham}. The theory is successful in the sense that it is able to reproduce some of the already known consequences attributed originally to a cosmological constant ($\Lambda$) when it is introduced inside the standard Einstein equations in General Relativity (GR). The theory is even able to reproduce the scale $r_V=(GM/m^2)^{1/3}$, known as the Vainshtein scale inside dRGT \cite{Vainshtein}. However such scale is not new and it has been part of the Einstein theory with cosmological constant for a long time if we analyze the local effects of $\Lambda$ \cite{My papers2, My papers2miau, HawkingBousso} and if additionally we relate the mass of the graviton to the $\Lambda$ scale. Recently some pathologies have been reported inside the dRGT formulation. Most of them related to superluminal propagation, matter-coupling inconsistencies, problems of causality, unitarity and predictability \cite{Deserpapers} and in addition, the impossibility of finding a consistent partially massless limit which would be useful at the moment of formulating a consistent bi-metric theory \cite{Deserpapers2}. At the cosmological level, some unstable solutions were found in \cite{Deserpapers3}. Some possible pathologies for the black-hole solutions were also reported in \cite{babichev, Kodama}. In fact, one of the most important predictions of Einstein gravity is the existence of black-holes. The black hole solution with a cosmological constant ($\Lambda$) is known as Schwarzschild de-Sitter (S-dS), which is stable in Einstein gravity. Any consistent modification of gravity should be able to reproduce the existence of stable black hole solutions as in the standard case. The black-hole solutions in dRGT were originally derived by Koyama et al. and subsequently by other authors \cite{Koyama}. Recently it was found that the S-dS black-hole solution in dRGT is stable, although some possible pathologies might remain due to the apparent degeneracy appearing through the time-components of the St\"uckelberg fields \cite{Kodama}. Inspired in this work, Babichev and colleagues have demonstrated that inside the bi-metric formulation of gravity, the Schwarzschild-like solution becomes stable if the fiducial metric is flat. Additionally some interesting properties for the solutions with one free-parameter were mentioned \cite{babichev}. By exploring the degeneracy arguments inside the S-dS solution in dRGT, the author commented that the usual notion of energy is not conserved in dRGT \cite{mine22}. The concept, can however be extended in order to write a conserved quantity under time-translations. In that sense, the theory is still predictable. The apparent lost of predictability already mentioned by Kodama and the author in \cite{Kodama}, only exists if a certain gauge transformation function $T_0(r,t)$ is not well defined at the background level, extending then the problem to the perturbative level. In the present manuscript, I go further with the definition of energy in dRGT by using the S-dS black-hole solution in order to obtain an expression for the Komar mass function. I demonstrate that the Komar mass function in dRGT depends on the dynamic of the St\"uckelberg fields through the gauge transformation function defined by $T_0(r,t)$. The standard definition of Komar mass function in GR is recovered as $\partial_rT_0(r,t)=0$. In such a case, $T_0(r,t)$ recovers the role of time. In some previous manuscript, it was demonstrated that the ADM mass is not conserved if we consider a linear massive term \cite{competition}. However, this is not necessarily the case at the non-linear level. It is true is that the usual notion of energy is not conserved, however the concept can be extended in order to include the dynamic of the St\"uckelberg fields. If the Killing vector is taken in the standard time-like direction, the effects of the St\"uckelberg fields will appear as spatial dependence of the gauge transformation function. In \cite{competition}, the space was considered to be asymptotically flat. This in principle can be done, but at the non-linear level it is not correct because in such a case we are covering scales larger than the cosmological horizon where there is no guarantee that our coordinate system is still the appropriate one. In fact we have to be careful in working inside the region where our static (quasi-stationary approach in dRGT in unitary gauge) coordinates are valid. The region remains inside the two horizons, namely, the black-hole horizon and the cosmological one if we consider the S-dS solutions. The present manuscript is organized as follows:
In Section (\ref{eq:dRGT}), I introduce the static S-dS black-hole solution in the standard theory of GR; in Section (\ref{eq:Final1}), I introduce the S-dS solution, but this time inside the dRGT massive gravity formulation in unitary gauge; 
in Section (\ref{eq:This la}), I derive the conserved quantities associated with the motion of a massive test particle. This part was already derived in \cite{My papers2miau, mine22}; in Section (\ref{eq:Komarenergy}), I derive the Komar energy function for the S-dS solution in dRGT. This part is the new contribution of this manuscript; finally in Section (\ref{eq:Conclude}), I conclude.

\section{The Schwarzschild de-Sitter space in General Relativity: Motion of a test particle}   \label{eq:dRGT}

The Schwarzschild-de Sitter metric in static coordinates, is given by:

\begin{equation}   \label{eq:Sdsm}
ds^2=-e^{\nu(r)}dt^2+e^{-\nu(r)}dr^2+r^2d\theta^2+r^2\sin^2\theta d\phi^2,
\end{equation}

with:

\begin{equation}   \label{eq:e}
e^{\nu(r)}=1-\frac{r_s}{r}-\frac{r^2}{3r_\varLambda^2},
\end{equation}

where $r_s=2GM$ is the gravitational radius and $r_\Lambda=\frac{1}{\sqrt{\Lambda}}$ defines the cosmological constant scale.

\section{The Schwarzschild de-Sitter solution in dRGT: Unitary gauge} \label{eq:Final1}

In dRGT massive gravity theory, the action is defined as:

\begin{equation}   \label{eq:b1}
S=\frac{1}{2\kappa^2}\int d^4x\sqrt{-g}(R+m^2U(g,\phi)),
\end{equation}

with the potential expansion for $U(g, \phi)$ given by:

\begin{equation}   \label{eq:b2}
U(g,\phi)=U_2+\alpha_3 U_3+\alpha_4U_4,
\end{equation}

where $\alpha_3$ and $\alpha_4$ correspond to the two-free parameters of the theory. Inside this formulation, some black-hole solutions corresponding to different metrics have been found in \cite{Koyama}. Additionally, in \cite{Kodama}, the S-dS solution in unitary gauge was derived for two different cases. The first one, corresponds to the family of solutions satisfying the condition $\beta=\alpha^2$, where $\beta$ and $\alpha$ are related to the two free-parameters $\alpha_3$ and $\alpha_4$ \cite{Kodama}. In such a case, the gauge transformation function $T_0(r,t)$ becomes arbitrary but obeying a symmetry. The second one, corresponds to the family of solutions with two-free parameters satisfying the condition $\beta\leq\alpha^2$ with the gauge transformation function $T_0(r,t)$ constrained. The explicit relations between the two free-parameters $\alpha_3$ and $\alpha_4$ with respect to $\beta$ and $\alpha$ are:

\begin{equation}
\alpha=1+\alpha_3,\;\;\;\;\;\beta=3(\alpha_3+4\alpha_4).
\end{equation}

All the spherically symmetric solutions obtained in \cite{Kodama}, can be generically expressed as:

\begin{equation}
ds^2=g_{tt}dt^2+g_{rr}dr^2+g_{rt}(drdt+dtdr)+r^2d\Omega_2^2,
\end{equation}
 
where:

\begin{equation}   \label{eq:drgt metric}
g_{tt}=-f(r)(\partial_tT_0(r,t))^2,\;\;\;\;\;g_{rr}=-f(r)(\partial_rT_0(r,t))^2+\frac{1}{f(r)},\;\;\;\;\;g_{tr}=-f(r)\partial_tT_0(r,t)\partial_rT_0(r,t),
\end{equation}

with $f(r)=1-\frac{2GM}{r}-\frac{1}{3}\Lambda r^2$. The metric (\ref{eq:drgt metric}), contains all the degrees of freedom (5 in total). It means that we are working in the unitary gauge. In such a case, the fiducial metric is just the Minkowskian one given explicitly as:

\begin{equation}
f_{\mu\nu}dx^\mu dx^\nu=-dt^2+\frac{dr^2}{S_0^2}+\frac{r^2}{S_0^2}(d\theta^2+r^2sin^2\theta),
\end{equation}
 
where $S_0=\frac{\alpha}{\alpha+1}$. The St\"uckleberg fields take the standard form defined in \cite{Kodama}.

\section{Conserved quantities for a test particle moving in dRGT}   \label{eq:This la}

Inside the dRGT formulation of massive gravity, in unitary gauge, the quantity:

\begin{equation}   \label{eq:Falcao}
g_{\mu \nu}U^\mu U^\nu=C,
\end{equation}

is a constant of motion. It represents the Lagrangian of a test particle moving around a source. If we want to analyze the other conserved quantities, then it is convenient to write eq. (\ref{eq:Falcao}) explicitly as:

\begin{equation}   \label{eq:Messi}
g_{tt}\left(\frac{dt}{d\tau}\right)^2+g_{rr}\left(\frac{dr}{d\tau}\right)^2+2g_{tr}\left(\frac{dr}{d\tau}\right)\left(\frac{dt}{d\tau}\right)+g_{\phi\phi}\left(\frac{d\phi}{d\tau}\right)^2=C,
\end{equation}

where I have omitted the zenithal angle represented by $\theta$ because we can fix it due to the spherical symmetry of the metric. If we assume the metric to be stationary, then the gauge-transformation function $T_0(r,t)$ is linear in time and then the components of the metric ($g_{\mu\nu}$) are time-independent. In such a case, from (\ref{eq:Messi}), we can find the equations of motion for $t$ and $\phi$ as:

\begin{equation}   \label{eq:CR7}
\frac{d}{d\tau}\left(g_{tt}\left(\frac{dt}{d\tau}\right)+g_{rt}\left(\frac{dr}{d\tau}\right)\right)=0,
\end{equation}

\begin{equation}   \label{eq:CR7}
\frac{d}{d\tau}\left(r^2\left(\frac{d\phi}{d\tau}\right)\right)=0.
\end{equation}

The second equation is just the conservation of the angular momentum. The first equation is related to the energy-conservation. In GR, the term $g_{rt}$ vanishes since we have gauge freedom for the dynamical metric in such a case. Inside dRGT however, any attempt for removing the $r-t$ component of the metric, just translates degrees of freedom from the dynamical metric to the fiducial one and the physical effects of the $g_{rt}$ component, would just be translated to the fiducial metric. From eq. (\ref{eq:CR7}), the total energy is not conserved in its original form, namely, $E=g_{tt}dt/d\tau$. Instead, the conserved quantity associated with translations in time is given by:

\begin{equation}   \label{eq:Ronaldo}
g_{tt}\left(\frac{dt}{d\tau}\right)+g_{rt}\left(\frac{dr}{d\tau}\right)=E_{dRGT},
\end{equation}

where the subindex dRGT suggests that this quantity should be recognized as an extended total energy inside dRGT. Eq. (\ref{eq:Ronaldo}) however, tells us that the total energy in its usual form is a velocity-dependent quantity. For different values of $dr/d\tau$, the value of $E=g_{tt}\left(\frac{dt}{d\tau}\right)$ changes. Then any attempt for describing the motion of a particle (or perturbation) by using the standard notion of energy reproduces a degeneracy. $E_{dRGT}$ can be considered as a conserved quantity along the direction $T_0(r.t)$. This does not mean that the concept of time-like Killing vector should change. In fact, we can still define the time-lime Killing vector in its usual form (in agreement with GR):

\begin{equation}   \label{eq:Killing vector}
K^\mu\to(1,0,0,0),
\end{equation}

as $r\to\infty$ in asymptotically flat spaces, and $r_s<<r<<r_\Lambda$ for asymptotically de-Sitter spaces. Here I will consider this second situation, where however, a correction due to a normalization factor must be taken into account for the time-like Killing vector \cite{HawkingBousso}. General relativity has different notions of energy. Here I will consider the Komar energy.

\section{The Komar energy: The case of S-dS solution in dRGT}   \label{eq:Komarenergy}

I will derive the Komar energy in dRGT. From the previous section, we can expect some modification of the energy expression to be conserved under time translations. In \cite{competition}, the analysis for the ADM energy was done inside the linear regime of massive gravity. Here I consider the full non-linear regime in order to calculate the Komar energy function. In the non-linear regime, the field equations are:

\begin{equation}   \label{eq:Eins}
G_{\mu\nu}=-m^2X_{\mu\nu},
\end{equation}

where $G_{\mu\nu}$ is the Einstein tensor and $X_{\mu\nu}$ is the energy-momentum tensor derived from the variation of the potential with respect to the dynamical metric. There is an extra equation obtained from the variation of the potential with respect to the St\"uckelberg fields. This equation, which corresponds to the dynamic of the St\"uckelberg fields, is equivalent to the Bianchi identity only in the unitary gauge. Once we work outside the unitary gauge, the dynamic of the St\"uckelberg fields is not necessarily related to the energy-momentum conservation. In General Relativity, there are different notions of energy. All of them have a common feature. They correspond to the symmetry under time-translations. In dRGT non-linear massive gravity, care must be taken at the moment of calculating the energy by using this concept. The reason is that the usual notion of time can be kept only in the unitary gauge. Once we work outside this gauge, we have to take the arrow of time in agreement with the value assumed by the time component of the St\"uckelberg field. This in fact is just an illusion since the usual notion of time remains unchanged in a physical sense. In GR, we can define the following conserved current:

\begin{equation}   \label{eq:Curr}
J_T^\mu=K_\nu T^{\mu\nu},
\end{equation}

where $K_\mu$ is the form defined as $g_{\mu\nu}K^\nu$, with $K^\mu$ representing the time-like Killing vector. $T^{\mu\nu}$ is the energy-momentum tensor. From this previous expression, we can define the following energy inside a two sphere:

\begin{equation}   \label{eq:curr2}
E_T=\int_{\sum} d^3x\sqrt{\gamma}n_\mu J_T^\mu,
\end{equation}

where $n^\mu$ corresponds to a unit vector normal to the space-like hypersurface. It is true that in dRGT we can in principle use the same set of equations in order to evaluate the previous quantities. However care must be taken in such a case because the non-diagonal term $g_{tr}$ of the dynamical metric provides an extra contribution to the integral. That contribution cannot be gauged away because it comes from the St\"uckelberg fields as has been explained before. In GR however, the current defined in (\ref{eq:Curr}) is not good because in general we study the space-time in vacuum ($T_{\mu\nu}=0$). It is true that in massive gravity we do not work exactly in vacuum because we have a contribution coming from $X_{\mu\nu}$ defined in (\ref{eq:Eins}). However, at the background level, this quantity behaves exactly as the cosmological constant for certain family of solutions \cite{Koyama, Kodama}. Anyway, the definition (\ref{eq:curr2}), does not take into account the energy coming from the gravitational field itself. Then it is not a good concept of energy. Instead, we can define the following current:

\begin{equation}   \label{eq:MiauPotter}
J_R=K_\nu R^{\mu\nu},
\end{equation}

which is conserved if we keep in mind that $K^\mu$ represents a Killing vector and as a consequence, the form $K_\mu$ satisfies the Killing equation. Additionally the geometry is unchanged in the direction of the Killing vector. The energy associated to the current (\ref{eq:MiauPotter}), is given by:

\begin{equation}   \label{eq:Miau2}
E_R=\frac{1}{4\pi G}\int_{\sum}d^3x\sqrt{\gamma}n_\mu J_R^\mu.
\end{equation}

Taking into account that the Killing vector satisfies the condition $\nabla_\mu\nabla_\nu K^\mu=K^\mu R_{\mu\nu}$, then we can write the current (\ref{eq:MiauPotter}) as a total derivative given by $J^\mu_R=\nabla_\nu(\nabla^\mu K^\nu)$. Then the Komar energy given by (\ref{eq:Miau2}) can be expressed as:

\begin{equation}   \label{eq:Komint}
E_R=\frac{1}{4\pi G}\int_{\partial \sum}d^2x\sqrt{\gamma}n_\mu\sigma_\nu\nabla^\mu K^\nu.
\end{equation}

Here $\partial\sum$ is the boundary and it corresponds to a two-sphere ($S^2$) at large spatial scales, it has a metric $\gamma_{ij}$ and an outward pointing unit vector $\sigma^\mu$. In addition, $K^\mu\to(1,0,0,0)$ inside a region satisfying $r_s<<r<<r_\Lambda$. More exactly, the time-like Killing vector should be normalized by using the same method used by Bousso and Hawking \cite{HawkingBousso}. In general the Killing vector can be written as:

\begin{equation}
K^\mu=\gamma_t\frac{\partial}{\partial t},
\end{equation}

where $\gamma_t$ is just a normalization factor. It would be $\gamma_t=1$ if we normalize the Killing vector as $K^2\to-1$ as $r\to\infty$ as it is normally done in an asymptotically flat space. In this case however, we are considering an asymptotically de-Sitter space and we want to guarantee the well behavior of our coordinate system. Then we have to normalize the time-like Killing vector as:

\begin{equation}   \label{eq:extr22}
\gamma_t=(-g_{00})^{-1/2}_{r=r_V}.
\end{equation}

Here $r_V$ corresponds to the Vainshtein scale defined in \cite{My papers2miau} as an extremal condition for the dynamical metric in unitary gauge. The scale $r_V$ can be derived after solving the equation:

\begin{equation}   \label{eq:extr}
dg_{\mu\nu}=\left(\frac{\partial g_{\mu\nu}}{\partial r}\right)_tdr+\left(\frac{\partial g_{\mu\nu}}{\partial t}\right)_rdt=0,
\end{equation}

when all the degrees of freedom are inside the dynamical metric. For stationary coordinates, the previous condition is just reduced to $f'(r)=0$, which is in agreement with the Bousso-Hawking definition of static observer condition in the S-dS space in GR. Some differences might appear however, when we consider time-evolving geometries, even at the stationary level. For evaluating the integral (\ref{eq:Komint}), we have to take into account the appropriate normalizations for $n^\mu$ and $\sigma^\mu$. In this case, even if the Killing vector normalization changes, it depends explicitly on the mass of the black-hole and then it will be a constant with respect to spatial derivatives. Then we can use the same procedure for evaluating the Komar integral as it is normally done for the case of asymptotically flat space. Before evaluating the integral (\ref{eq:Komint}), we have to know explicitly the term $n_\mu\sigma_\nu\nabla^\mu K^\nu$. First, we can normalize $g_{\mu\nu}n^\mu n^\nu=-1$ and $g_{\mu\nu}\sigma^\mu\sigma^\nu=1$. It then provides the following results:

\begin{equation}
n_0=-\sqrt{f(r)}\partial_tT_0(r,t),\;\;\;\;\;\sigma_r=\sqrt{-f(r)(\partial_rT_0(r,t))^2+\frac{1}{f(r)}}.
\end{equation}

Then we get:

\begin{multline}    \label{eq:th}
n_\mu\sigma_\nu\nabla^\mu K^\nu=\frac{1}{\partial_tT_0(r,t)f(r)}\left(-f(r)^2(\partial_rT_0(r,t))^2+1\right)^{3/2}\Gamma^r_{00}\gamma_t\\
+f(r)\partial_rT_0(r,t)\left(\sqrt{-f(r)^2(\partial_rT_0(r,t))^2+1}\right)\Gamma^r_{r0}\gamma_t,
\end{multline}

where $\Gamma^\mu_{\nu\sigma}$ corresponds to the connection components. What remains to do is to evaluate the Christoffel connections. Here I will consider stationary backgrounds. Then the metric components are expected to be time-independent. In such a case, we expect $T_0(r,t)$ to be linear with respect to the standard time $t$, but arbitrary with respect to the coordinate $r$. Explicitly it would be of the form:

\begin{equation}
T_0(r,t)\backsim t+A(r),
\end{equation}

with $A(r)$ being an arbitrary function. Then it is clear that $\partial_t\partial_r T_0(r,t)=0$ and that $\partial_t T_0(r,t)\backsim 1$. The relevant Christoffel components are then given by:

\begin{equation}
\Gamma^r_{00}=\frac{1}{2}(\partial_tT_0(r,t))^2f(r)f'(r),\;\;\;\;\;\Gamma^r_{r0}=\frac{1}{2}(\partial_rT_0(r,t))(\partial_tT_0(r,t))f(r)f'(r).
\end{equation}

If we replace these results in (\ref{eq:th}). Then we get:

\begin{equation}    \label{eq:th2}
n_\mu\sigma_\nu\nabla^\mu K^\nu=\frac{\gamma_t}{2}\partial_tT_0(r,t)f'(r)\sqrt{-f(r)^2(\partial_rT_0(r,t))^2+1}.
\end{equation}

The determinant of the induced metric is given by:

\begin{equation}   \label{eq:th3}
\sqrt{\gamma^{(2)}}=r^2sin\theta.
\end{equation}

If we introduce the results (\ref{eq:th2}) and (\ref{eq:th3}) inside (\ref{eq:Komint}), then we get:

\begin{equation}   \label{eq:MdRGT}
E_R=\frac{\gamma_t}{2G}\partial_tT_0(r,t)r^2f'(r)\sqrt{-f(r)^2(\partial_rT_0(r,t))^2+1}=M_{dRGT}.
\end{equation}

This previous result should be considered as the Komar mass function in dRGT massive gravity. Note that the value of the mass will in general depend on the gauge transformation function which contains the information of the dynamic of the St\"uckelberg fields. As $\partial_rT_0(r,t)=0$, we recover the mass function already behaving like $E_R\backsim M-\Lambda r^3/6$ if we take the appropriate sign convention for $\Lambda$ \cite{Final} as in GR. Note that the Komar mass function given by (\ref{eq:MdRGT}) is a well defined quantity only if the following condition is satisfied:

\begin{equation}   \label{eq:amazing}
-\int\frac{dr}{f(r)}+C_1(t)<T_0(r,t)<\int\frac{dr}{f(r)}+C_2(t),
\end{equation}

where $C_{1,2}(t)$ are time-dependent functions. Since we are considering here only stationary backgrounds, then $C(t)\backsim t$, as has been mentioned before. Note that the condition (\ref{eq:amazing}) is satisfied by the second solution obtained in \cite{Kodama}. That solution corresponds to the case with two-free parameters with the gauge transformation function defined as:

\begin{equation}
T_0(r,t)=St\pm\int^{Sr}\left(\frac{1}{f(u)}-1\right).
\end{equation}

On the other hand, for the case $\alpha=\beta^2$, since $T_0(r,t)$ is in principle arbitrary (although related to a symmetry), then it is important to keep in mind the condition (\ref{eq:amazing}). Even if we ignore the $\Lambda$ contribution in (\ref{eq:MdRGT}), the result is different with respect to the conserved quantity obtained from GR. This is expected because we have the energy contribution of the extra-degrees of freedom activated after the Vainshtein radius through the spatial dependence of the gauge transformation function. In fact, $M_{dRGT}$ would depend on the dynamics of the St\"uckelberg fields. $M_{dRGT}$ becomes a conserved quantity (independent on r), if:

\begin{equation}
\pm f(r)\partial_rT_0(r,t)=Constant<1.
\end{equation}

It can be also considered a conserved quantity (in some sense) if, independent on the explicit solution for $T_0(r,t)$, the following attractor condition is satisfied:

\begin{equation}   \label{eq:Withlanb}
\pm f(r)\partial_rT_{0}(r,t)_{r>>r_s}\to 0,
\end{equation}

or constant smaller than unity. In eq. (\ref{eq:Withlanb}), we can forget for a while the $\Lambda$ contribution which will appear anyway for large $r$. In such a case, $T_0(r,t)$ would correspond to an attractor solution for the dynamic involved.

\section{Conclusions}   \label{eq:Conclude}

I have obtained a compact expression of the Komar mass function for the case of a stationary background inside the dRGT non-linear formulation of massive gravity. The expression corresponds to the S-dS black-hole solution found in \cite{Kodama}. In general, the quantity $M_{dRGT}$ is conserved under some conditions depending on the dynamics of the St\"uckelberg fields. $M_{dRGT}$ represents the contributions coming from all the degrees of freedom, including the St\"uckelberg fields. In the special case where $\partial_rT_0(r,t)=0$, the results are reduced to the standard Komar/ADM mass energy as defined in General Relativity (GR).\\\\

{\bf Acknowledgement}
The author would like to thank Masahide Yamaguchi and Masaru Siino from Tokyo Institute of Technology for useful discussions around this topic, as well as Yen Chin Ong from National Taiwan University for the suggested reference \cite{competition}.

\end{document}